\documentclass[prl,twocolumn,aps,showpacs]{revtex4}
\usepackage{epsfig}
\usepackage{bm}
\begin{document}
\date{\today}

\title{Trapped fermions with density imbalance in the BEC limit }
\author{P. Pieri and G.C. Strinati}
\affiliation{Dipartimento di Fisica Universit\`{a} di Camerino, I-62032 
Camerino, Italy}

\begin{abstract}
We analyze the effects of imbalancing the populations of two-component trapped fermions, in the BEC
limit of the attractive interaction between different fermions.
Starting from the gap equation with two fermionic chemical potentials, we derive a set of coupled equations
that describe composite bosons and excess fermions.
We include in these equations the processes leading to the correct dimer-dimer and dimer-fermion
scattering lengths.
The coupled equations are then solved in the Thomas-Fermi approximation to obtain the density profiles for 
composite bosons and excess fermions, which are relevant to the recent experiments with trapped fermionic atoms.
\end{abstract}

\maketitle

Lately there has been mounting interest in trapped atomic fermion superfluids, both experimentally
and theoretically. These systems allows one to explore the BCS-BEC crossover with the use of Fano-Feshbach
resonances.
Interesting many-body issues for fermionic superfluids have been addressed in this context, aiming also at making
connections with high-temperature superconductors and nuclear and quark matters.
Until recently, only the case of two fermionic species with equal populations was experimentally explored.

Two new experimental studies of fermionic superfluidity with trapped atoms having imbalanced populations
\cite{Ketterle-d-i,Hulet-d-i} have raised novel interest in these systems.
Density profiles of the two fermionic species \cite{Ketterle-d-i,Hulet-d-i} as well as vortices \cite{Ketterle-d-i}
have been detected.
A quantum phase transition to the normal state on the weak-coupling side of the BCS-BEC crossover \cite{Ketterle-d-i}
as well as a phase separation in the crossover region \cite{Hulet-d-i,Ketterle-d-i}, both driven by density imbalance, have
been identified.

The many-body problem becomes reacher when the fermionic populations are imbalanced, since this additional degree of
freedom gives rise to new phenomena.
The effects of density imbalance on fermionic superfluids have been studied theoretically mostly within mean field and
on the weak-coupling (BCS) side of the crossover, both for the homogeneous \cite{BCS-d-i-homogeneous} and trapped
case \cite{BCS-d-i-trapped}.
Only recently these calculations have been extended to cover the BCS-BEC crossover \cite{BCS-BEC-d-i-homogeneous}, but without considering the
effects of the trap which are essential to account for the experimental results
with density imbalance \cite{Ketterle-d-i,Hulet-d-i}.

The effects of the trap are especially important to produce a phase separation between the density profiles of the two
fermionic species, which may otherwise remain at the verge of an instability for a homogeneous system.
In this respect, consideration of the strong-coupling (BEC) side of the crossover appears relevant, since phase
separation is more robust in this limit.
[On the weak-coupling side, on the other hand, density imbalance acts to quickly destroy the superfluid phase.]
In addition, on the BEC side theoretical studies can rely on more accurate treatments beyond mean field, by exploiting
the diluteness condition of the system.
From previous experience on the BCS-BEC crossover, one expects the results obtained on the BEC side to be
qualitatively similar to those occurring near unitarity.

Accordingly, in this paper we analyze the BEC side of the crossover where trapped fermions with imbalanced
populations get rearranged into a system of composite bosons (dimers) plus a number of excess (unpaired) fermions.
We consider specifically the limit of low temperature, whereby all dimers are condensed, and derive a set of coupled
equations describing the dimers and the excess fermions, which are interacting through dimer-dimer and dimer-fermion
scattering processes.
To the extent that the system is dilute, we can consider these processes exhaustively.
Due to these interactions, the mutual effects between dimers and excess fermions need to be dealt with self-consistently
in the equations.
This results in a nontrivial evolution of their density profiles as the degree of density imbalance increases.
We determine numerically these density profiles within a local approximation and find that
dimers and excess fermions tend to reside in different spatial regions of the trap.
We also show how from these profiles one can, in reverse, extract the ratio $a_{{\rm BF}}/a_{{\rm F}}$ of the dimer-fermion to the
fermion-fermion scattering lengths.

We first consider the derivation of the above coupled equations within mean field.
When the populations of the two fermionic species (labeled by $\uparrow$ and $\downarrow$) are equal, it has already been
shown \cite{PS-2003} that the Bogoliubov-de Gennes (BdG) equations for trapped fermions can be mapped in the BEC
limit onto the Gross-Pitaevskii equation for composite bosons, the only remnant of the original fermionic nature residing
in the dimer-dimer scattering length entering that equation.
In the case of interest here of imbalanced populations, this mapping has to be reconsidered since 
composite bosons and excess fermions coexist and mutually interact.
In the following, we shall recall schematically the derivation of Ref.\cite{PS-2003}, and indicate where and how it needs
to be modified to account for population imbalance.

The presence of two fermionic species with populations $N_{\uparrow}$ and $N_{\downarrow}$ requires us to
introduce two chemical potentials $\mu_{\uparrow}$ and $\mu_{\downarrow}$.
(For definiteness, we assume $N_{\uparrow} \ge N_{\downarrow}$.)
In analogy to Ref.\cite{PS-2003}, we consider the noninteracting Green's functions 
which satisfy the equation
$\left[ i \omega_{s} \, - \, \mathcal{H}_{\sigma}(\mathbf{r}) \right] \,
\tilde{\mathcal{G}_{0}}(\mathbf{r},\mathbf{r}';\omega_{s}|\mu_{\sigma}) \, = \, \delta(\mathbf{r} - \mathbf{r}')$,
being subject to the trap potential $V(\mathbf{r})$ entering the single-particle Hamiltonian
$\mathcal{H}_{\sigma}(\mathbf{r}) = - \nabla^{2} /(2m) + V(\mathbf{r})  -  \mu_{\sigma}$.
Here, $\sigma= (\uparrow,\downarrow)$, $m$ is the fermion mass, and $\omega_{s} = (2s+1)\pi/\beta$ ($s$ integer) is a
fermionic Matsubara frequency ($\beta$ being the inverse temperature).
With these noninteracting Green's functions we form the matrix:
$$
\hat{\mathcal{G}_{0}}(\mathbf{r},\mathbf{r}';\omega_{s}) =
\left[ \begin{array}{cc}
\tilde{\mathcal{G}_{0}}(\mathbf{r},\mathbf{r}';\omega_{s}|\mu_{\uparrow}) & 0 \\
0 & - \tilde{\mathcal{G}_{0}}(\mathbf{r}',\mathbf{r};-\omega_{s}|\mu_{\downarrow}) \end{array}
\right] \, .           
$$
The corresponding interacting single-particle Green's functions within mean field are obtained by solving the
integral equation:
\begin{eqnarray}
& &\hat{\mathcal{G}}(\mathbf{r},\mathbf{r}';\omega_{s})= \hat{\mathcal{G}_{0}}(\mathbf{r},\mathbf{r}';\omega_{s})\nonumber\\
& & \phantom{1111}+\int \! d\mathbf{r}'' \, \hat{\mathcal{G}_{0}}(\mathbf{r},\mathbf{r}'';\omega_{s}) \hat{{\rm B}}(\mathbf{r}'')
\hat{\mathcal{G}}(\mathbf{r}'',\mathbf{r}';\omega_{s}) \label{integral-equation}
\end{eqnarray}
where
$$
\hat{{\rm B}}(\mathbf{r}) =
\left[ \begin{array}{cc}
           0 & \Delta(\mathbf{r}) \\
\Delta^{*}(\mathbf{r}) & 0 \end{array}
\right]            \label{B-matrix}
$$
contains the local gap $\Delta(\mathbf{r})$.
Solution of the integral equation (\ref{integral-equation}) is equivalent to the solution of the BdG
equations in the trap in the presence of density imbalance.

The elements $(1,1)$ and $(2,2)$ of the matrix (\ref{integral-equation}) determine the local densities, while
the element $(2,1)$ determines the local gap.
To obtain the coupled equations of interest we expand them up to order $\Delta^{2}$ and $\Delta^{3}$, respectively, in
analogy to what was done in Ref.\cite{PS-2003} in the absence of density imbalance.
The gap equation reads:
\begin{eqnarray}
0 & = & \frac{1}{v_{0}} \Delta^{*}(\mathbf{r}) - 
        \frac{1}{\beta} \sum_{s} e^{-i\omega_{s} \eta}  G_{21}(\mathbf{r},\mathbf{r};\omega_{s}) \nonumber\\
& \cong & \frac{1}{2} b_{0}(\mathbf{r}) \nabla^{2} \Delta^{*}(\mathbf{r}) + 
\left( \frac{1}{v_{0}} +  a_{0}(\mathbf{r}) \right) \Delta^{*}(\mathbf{r})\nonumber\\  
&\phantom{111}& +\, c_{0}(\mathbf{r}) |\Delta(\mathbf{r})|^{2} \Delta^{*}(\mathbf{r})\label{gap-equation}  
\end{eqnarray}
where $\eta$ is a positive infinitesimal and $v_{0} \, (<0)$ is the strength of the attractive contact potential between fermions of opposite spins.
As in Ref.\cite{PS-2003}, the coefficients $a_{0}$, $b_{0}$, and $c_{0}$ of Eq.(\ref{gap-equation}) are expressed in terms
of the noninteracting Green's functions $\tilde{\mathcal{G}_{0}}(\mu_{\sigma})$, here with chemical potentials
$\mu_{\sigma}$ appropriate to imbalanced populations.
While $b_{0}(\mathbf{r}) \cong m a_{{\rm F}}/(16 \pi)$ retains the same value as for equal populations, $a_{0}$ and $c_{0}$ are 
affected by density imbalance in a relevant way.
With the local \emph{ansatz} for $\tilde{\mathcal{G}_{0}}(\mu_{\sigma})$ introduced in Ref.\cite{PS-2003} in terms
of the noninteracting Green's functions of the associated homogeneous problem, we obtain at low temperature:
\begin{eqnarray}
&&\frac{1}{v_{0}} + a_{0}(\mathbf{r}) =  \frac{1}{v_{0}} + \int \! \frac{d\mathbf{k}}{(2 \pi)^{3}}\nonumber\\
&& \times
\frac{\left[1 - f \left(\frac{\mathbf{k}^{2}}{2m} + V(\mathbf{r}) - \mu_{\uparrow} \right)
               - f \left(\frac{\mathbf{k}^{2}}{2m} + V(\mathbf{r}) - \mu_{\downarrow} \right)\right]}{\frac{\mathbf{k}^{2}}{m} 
- \mu_{\uparrow} - \mu_{\downarrow} + 2 V(\mathbf{r})} \nonumber \\
& &\cong \frac{m^{2} a_{{\rm F}}}{8 \pi} \left[ \mu_{{\rm B}} - 2 V(\mathbf{r}) \right]
                - \frac{1}{\epsilon_{0}} \,\delta n_{f}(\mathbf{r}) \label{approximate-a-0}
\end{eqnarray}
where $f(\epsilon)= \left( \exp (\beta \epsilon) + 1 \right)^{-1}$ is the Fermi function.
We have made use of the fermion bound-state equation with binding energy
$\epsilon_{0}=(m a_{{\rm F}}^{2})^{-1}$, introduced the bosonic chemical potential
$\mu_{{\rm B}} = \mu_{\uparrow} + \mu_{\downarrow} + \epsilon_{0}$, and anticipated that $\mu_{\downarrow} \sim -\epsilon_{0}$
while $\mu_{\uparrow} \sim \epsilon_{{\rm F}}(\delta n)$ where $\epsilon_{{\rm F}}(\delta n)$ is the Fermi level of \emph{free\/}
excess fermions with density:
\begin{equation}
\delta n_{f}(\mathbf{r}) = \int \! \, \frac{d\mathbf{k}}{(2 \pi)^{3}} \,
       f \left(\frac{\mathbf{k}^{2}}{2m} + V(\mathbf{r}) - \mu_{\uparrow} \right) \, .             \label{free-density}
\end{equation}
Equation (\ref{approximate-a-0}) can be cast in the meaningful form
\begin{equation}
\frac{1}{v_{0}} + a_{0}(\mathbf{r}) \, \cong \, \frac{m^{2} a_{{\rm F}}}{8 \pi}
\left[ \mu_{{\rm B}}  - 2 V(\mathbf{r}) \, - \, \frac{3 \pi a_{{\rm BF}}}{m}  \delta n_{f}(\mathbf{r}) \right] \, ,
                              \label{approximate-a-0-final}
\end{equation}
where we have identified the dimer-fermion scattering length $a_{{\rm BF}}$ which takes the Born value $8a_{{\rm F}}/3$ at the
mean-field level.
We obtain in addition:
\begin{equation}
c_{0}(\mathbf{r}) = 
- \frac{\partial }{\partial \mu_{\uparrow}} \frac{\partial }{\partial \mu_{\downarrow}} a_{0}(\mathbf{r})
 \cong  - \frac{m^{3} a_{{\rm F}}^{3}}{16 \pi} +  m^{2} a_{{\rm F}}^{4} 
\frac{\partial \delta n_{f}(\mathbf{r})}{\partial \mu_{\uparrow}} \,. \label{approximate-c-0-final}
\end{equation}
The gap equation (\ref{gap-equation}) eventually reduces to the following equation for the condensate wave function
$\Phi(\mathbf{r}) = \sqrt{m^{2} a_{{\rm F}}/(8 \pi)} \, \Delta(\mathbf{r})$:
\begin{eqnarray}
- \frac{\nabla^{2}}{2 m_{{\rm B}}} \Phi(\mathbf{r})
&+& \left[ 2 V(\mathbf{r}) + \frac{3 \pi a_{{\rm BF}}}{m} \delta n(\mathbf{r}) \right] \Phi(\mathbf{r})\nonumber\\
&+& \frac{4  \pi a_{{\rm B}}}{m_{{\rm B}}}  |\Phi(\mathbf{r})|^{2} \Phi(\mathbf{r})
= \mu_{{\rm B}} \Phi(\mathbf{r})  \label{gap-equation-final}
\end{eqnarray}
where $m_{{\rm B}}=2m$ is the dimer mass and the dimer-dimer scattering length $a_{{\rm B}}$ takes the Born value $2a_{{\rm F}}$
at the mean-field level.
Note how the last contribution in Eq.(\ref{approximate-c-0-final}) combines with the term proportional to $\delta n_{f}$
in Eq.(\ref{approximate-a-0-final}), to yield in Eq.(\ref{gap-equation-final}) the \emph{density of excess fermions}:
\begin{eqnarray}
&&\delta n(\mathbf{r})=  \delta n_{f}(\mathbf{r}) - \frac{3 \pi a_{{\rm BF}}}{m}  |\Phi(\mathbf{r})|^{2} 
\frac{\partial \delta n_{f}((\mathbf{r})}{\partial \mu_{\uparrow}}\label{interacting-density}\\
 && \cong  \int \!\! \, \frac{d\mathbf{k}}{(2 \pi)^{3}} \,
f\left(\frac{\mathbf{k}^{2}}{2m}+V(\mathbf{r})+\frac{3\pi a_{{\rm BF}}}{m} |\Phi(\mathbf{r})|^{2}-\mu_{\uparrow}\right) 
 \nonumber . 
\end{eqnarray}
By this mechanism, Eq.(\ref{gap-equation-final}) contains the effect of the fermionic distribution $\delta n(\mathbf{r})$
on the bosonic one, an effect which is reciprocated in Eq.(\ref{interacting-density}) by the bosonic distribution
$|\Phi(\mathbf{r})|^{2}$ on the fermionic one.

The final equation results from considering the total density $n(\mathbf{r})= n_{\uparrow}(\mathbf{r}) + n_{\downarrow}(\mathbf{r})$ where:
\begin{eqnarray}
n(\mathbf{r})&=&\frac{1}{\beta} \sum_{s} \left [e^{i\omega_{s} \eta} \, G_{11}(\mathbf{r},\mathbf{r};\omega_{s}) -
e^{-i\omega_{s} \eta} \, G_{22}(\mathbf{r},\mathbf{r};\omega_{s}) \right]\nonumber \\
& \cong & \delta n(\mathbf{r}) + 2 \, |\Phi(\mathbf{r})|^{2} \,\, . \label{density-equation-final}
\end{eqnarray}

Equations (\ref{gap-equation-final}) and (\ref{density-equation-final}), together with the definition
(\ref{interacting-density}), are the desired closed set of equations which determine $\Phi(\mathbf{r})$ and $\delta n(\mathbf{r})$
in the BEC limit, for given total number of fermions $N_{\uparrow} + N_{\downarrow} = \int \! d\mathbf{r} \, n(\mathbf{r})$
and density imbalance $N_{\uparrow} - N_{\downarrow} = \int \! d\mathbf{r} \, \delta n(\mathbf{r})$.

The validity of these equations can be extended by improving on the values of the scattering lengths $a_{{\rm BF}}$
and $a_{{\rm B}}$.
At the mean-field level so far considered, these values correspond to the Born approximation, yielding
$a_{{\rm BF}}=8 a_{{\rm F}}/3$ and $a_{{\rm B}}=2 a_{{\rm F}}$.
The exact value $1.18 a_{{\rm F}}$ of $a_{{\rm BF}}$ has been known for some time \cite{STM-1957}, while the exact value $0.6 a_{{\rm F}}$
of $a_{{\rm B}}$ was determined only recently \cite{PSS-2004}.
These values have also been obtained by diagrammatic methods in the limit of vanishing density, for $a_{{\rm BF}}$
in Ref.\cite{BK-1998} and for $a_{{\rm B}}$ in Ref.\cite{KC-2005}.
To improve on the derivation of Eqs.(\ref{gap-equation-final}) and (\ref{density-equation-final}), so as to include
the exact values of $a_{{\rm BF}}$ and $a_{{\rm B}}$, we thus need to identify additional fermionic diagrams beyond mean field,
which contain the correct diagrammatic sequences for $a_{{\rm BF}}$ and $a_{{\rm B}}$ as sub-units.
This can be achieved as follows.
\begin{figure}
\begin{center}
\epsfxsize=7cm
\epsfbox{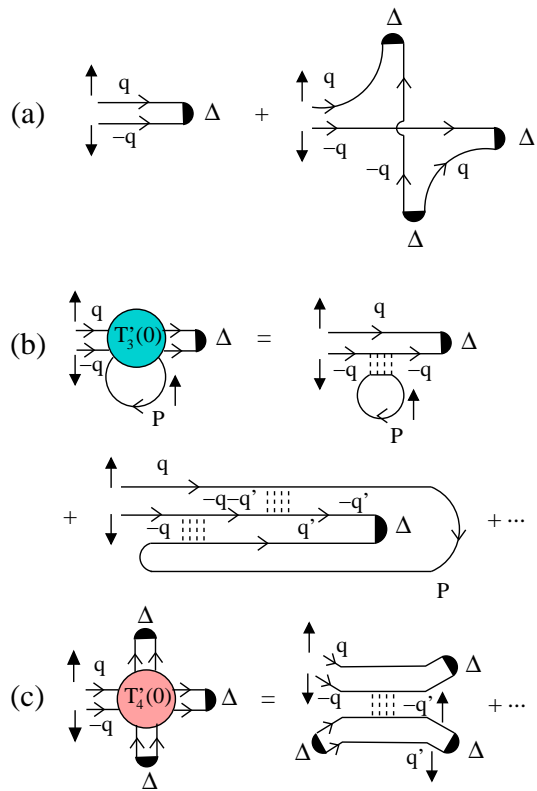}
\vspace{0.5cm}
\caption{Tadpole diagrams for composite bosons representing the gap equation in the BEC limit:
                   Mean-field contributions (a); Contributions including the dimer-fermion (b) and dimer-dimer
                   (c) scattering processed beyond the Born approximation.
                   Full line: fermionic propagator of given spin; broken line: fermionic interaction $v_{0}$.} 
\end{center}
\end{figure}

In the BEC limit, the gap equation can be interpreted as the condition of vanishing ``tadpole'' insertions for
composite bosons, with a natural extension of what is done for point-like bosons \cite{Popov}.
Within mean field, the diagrams representing this condition are depicted in Fig.1(a).
Here, a composite-boson propagator with zero four-momentum can be inserted from the left, while the gap $\Delta$
corresponds to a condensate line.
We have already shown that these diagrams contain the dimer-fermion and dimer-dimer scattering processes within
the Born approximation.
Additional tadpole diagrams, whose effect is to modify the values of $a_{{\rm BF}}$ and $a_{{\rm B}}$ in
Eq.(\ref{gap-equation-final}), are depicted in Figs.1(b) and 1(c), in the order.
These diagrams exclude the Born contributions already included in the mean-field diagrams of Fig.1(a).
A comment is in order on how to factor out from the diagrams of Fig.1(b) the product $a_{{\rm BF}}$ times $\delta n$
that enters Eq.(\ref{gap-equation-final}).
From the structure of these diagrams one concludes that the integration over the wave vector $\mathbf{P}$ is
bounded within the Fermi sphere with radius $\sqrt{2 m \mu_{\uparrow}}$, while the remaining integrations over the wave
vectors $\mathbf{q}$, $\mathbf{q'}$, $\cdots$, extend outside this Fermi sphere.
Accordingly, one may neglect the $P$-dependence everywhere in the diagrams of Fig.1(b), \emph{except\/} in the
fermion propagator labeled by $P$ and corresponding to a spin-$\uparrow$ fermion.
The density of excess fermions results in this way from the $P$-integration \cite{footnote-2}, while the remaining parts
of the diagrams yield the exact dimer-fermion scattering matrix $T^{'}_{3}(0)$,
which excludes the Born contribution resulting from mean field.

A similar analysis can be carried out for the density equation (\ref{density-equation-final}), where only the value of the dimer-fermion scattering
length $a_{{\rm BF}}$ needs to be corrected beyond mean field [cf. Eq.(\ref{interacting-density})].
In this case the relevant diagrams are similar to those of Fig.1(b), but with an additional condensate ($\Delta$)
line in the place of the incoming composite-boson propagator with zero four-momentum and with a doubling of the
spin-$\uparrow$ fermion propagator with four-momentum $P$.
This doubling provides the derivative with respect to $\mu_{\uparrow}$ in Eq.(\ref{interacting-density}), while the
remaining parts of the diagrams yield again $T^{'}_{3}(0)$.

We thus solve the coupled equations (\ref{gap-equation-final}) and (\ref{density-equation-final}) with the exact values
of $a_{{\rm BF}}$ and $a_{{\rm B}}$.
To make contact with the recent experimental findings
\cite{Ketterle-d-i,Hulet-d-i}, we determine the density profiles $\delta n({\bf r})$ and $n_0({\bf r})=|\Phi({\bf r})|^2$  by 
solving these equations  in the Thomas-Fermi approximation for
a spherical trap, as functions of the \emph{asymmetry parameter\/}
$\alpha=(N_{\uparrow} - N_{\downarrow})/(N_{\uparrow} + N_{\downarrow})$ (with $0 \le \alpha \le 1$).
\begin{figure}
\begin{center}
\epsfxsize=6cm
\epsfbox{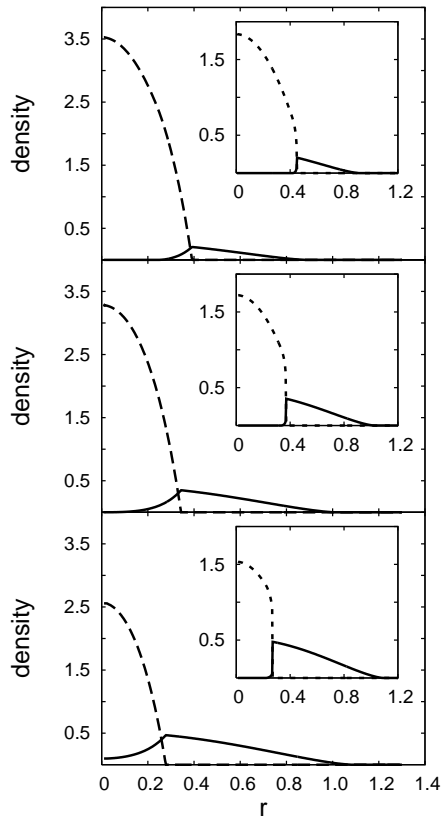}
\vspace{0.5cm}
\caption{Radial density profiles $\delta n(r)$ (full lines) and $n_{0}(r)$ (broken lines) vs
                   $r=|\mathbf{r}|$ for $(k_{{\rm F}} a_{{\rm F}})^{-1}=3$ and $\alpha = 0.2$ (upper panel), $\alpha = 0.5$
                   (middle panel), $\alpha = 0.8$ (lower panel)
                   [$r$ is in units of $R_{{\rm TF}}$ and densities are in units of $(N_{\uparrow} + N_{\downarrow}) / R_{{\rm TF}}^{3}$].
                   The insets show the results for $(k_{{\rm F}} a_{{\rm F}})^{-1}=1$ at the same values of $\alpha$.
                   Here, $R_{{\rm TF}}$ and $k_{{\rm F}}$ are the TF radius and Fermi wave vector for equal populations.} 
\end{center}
\end{figure}

In Fig.2 we report the radial density profiles $\delta n(r)$ and $n_{0}(r)$ vs the distance $r=|\mathbf{r}|$
from the center of the trap, for three characteristic values of $\alpha$ and for the coupling
$(k_{{\rm F}} a_{{\rm F}})^{-1}=3$ on the BEC side of the crossover.
The spatial separation between the condensed composite bosons and the excess fermions is evident from these plots.
For the coupling here considered, this phase separation occurs for all values of $\alpha$, because $\delta n(r)$ tends 
to set outside the region occupied by $n_{0}(r)$.
Note that, for each value of $\alpha$, the maximum of $\delta n(r)$ occurs where $n_{0}(r)$ vanishes.
Note also the progressive size shrinking of $n_{0}(r)$ for increasing $\alpha$, with a simultaneous penetration of
$\delta n(r)$ toward the center of the trap.
The insets show the results for the smaller coupling $(k_{{\rm F}} a_{{\rm F}})^{-1}=1$.
The phase separation between the two species appears sharper in this case, corresponding to enhanced effects of the
dimer-fermion repulsion.
For the column density profiles reported in Fig.3 (obtained from the above radial profiles by integrating over, say,
the $z$ axis) the phase separation appears less pronounced even for $(k_{{\rm F}} a_{{\rm F}})^{-1}=1$, since the
corresponding distribution $\int \! dz \, \delta n(\mathbf{\rho},z)$ leaks toward $\mathbf{\rho}=0$ where it acquires
a finite value.
In particular, for the coupling $(k_{{\rm F}} a_{{\rm F}})^{-1}=1$ near the boundary of the BEC region our column density
profiles are similar to those reported in Fig.2 of Ref.\cite{Hulet-d-i}.
To enhance the visibility of phase separation, it would therefore be preferable to consider radial rather than column
density profiles.
\begin{figure}
\begin{center}
\epsfxsize=6cm
\epsfbox{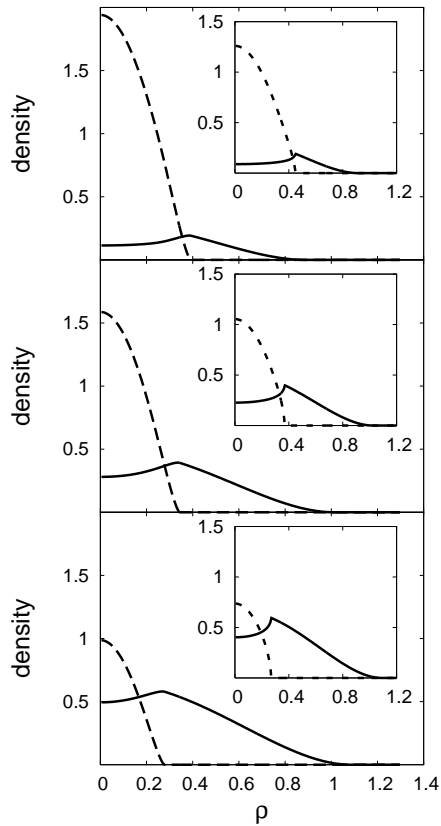}
\vspace{0.5cm}
\caption{Column density profiles $\delta n(\rho)$ (full lines) and $n_{0}(\rho)$ (broken lines) vs $\rho$
                    for $(k_{{\rm F}} a_{{\rm F}})^{-1}=3$ and $\alpha = 0.2$ (upper panel), $\alpha = 0.5$
                   (middle panel), $\alpha = 0.8$ (lower panel)
                   [$\rho$ is in units of $R_{{\rm TF}}$ and densities are in units of $(N_{\uparrow} + N_{\downarrow}) / R_{{\rm TF}}^{2}$].
                   The insets show the results for $(k_{{\rm F}} a_{{\rm F}})^{-1}=1$ at the same values of $\alpha$.
                   Here, $R_{{\rm TF}}$ and $k_{{\rm F}}$ are the TF radius and Fermi wave vector for equal populations.} 
\end{center}
\end{figure}

Our calculation does not reveal evidence for a critical value $\alpha_{c}$ below which the phase separation does not occur.
This somewhat contrasts with the experimental claim made in Ref.\cite{Hulet-d-i}, where at unitarity $\alpha_{c}$
was estimated to be about $0.10$.
This difference may either be due to the limited experimental resolution when revealing small
excess densities, or possibly to the vanishing of $\alpha_{c}$ when the
BEC region (considered in our calculation) is approached.
\begin{figure}
\begin{center}
\epsfxsize=7cm
\epsfbox{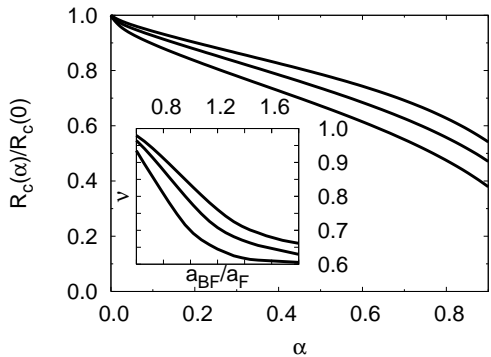}
\caption{Critical radius $R_{{\rm c}}$ of the condensate density vs the asymmetry parameter $\alpha$ (normalized to the
                   value $R_{{\rm c}}(0)$ at $\alpha=0$) for the coupling values $(k_{{\rm F}} a_{{\rm F}})^{-1}=(1,2,3)$ from bottom to top.
                   The inset shows the ratio $\nu$ between the column density of excess fermions at the center of the trap and its
                   maximum value at $R_{{\rm c}}$ vs $a_{{\rm BF}}/a_{{\rm F}}$, for $\alpha=0.5$ and the coupling values
                   $(k_{{\rm F}} a_{{\rm F}})^{-1}=(2,3,4)$ from bottom to top.} 
\end{center}
\end{figure}

In Fig.4 the critical radius $R_{{\rm c}}$ of the condensate density is plotted vs $\alpha$ for different
couplings, showing specifically how the condensate density shrinks for varying $\alpha$ and coupling.
This quantity should have direct experimental access, because it identifies also the position of the maximum of the density of excess fermions.
This maximum value can be compared with the value of the excess density at the center of the trap, to determine 
how their ratio $\nu$ depends on the dimer-fermion scattering length $a_{{\rm BF}}$ (assuming that the value of
$a_{{\rm B}}$ is independently determined).
Column density profiles are found to have a more marked dependence on $a_{{\rm BF}}$ than radial density profiles.
Accordingly, in the inset of Fig.4 we report the corresponding ratio $\nu$ vs $a_{{\rm BF}}$ for $\alpha=0.5$ and different couplings.
The rather marked dependence on $a_{{\rm BF}}$ shown by this quantity should make it possible to extract from the experimental
data the expected value $1.18$ of $a_{{\rm BF}}/a_{{\rm F}}$, using our plots for calibration.

We mention, finally, that, to compare with Fig.2D of Ref.\cite{Hulet-d-i}, we have also calculated the ratio $\nu$ for
$\alpha=0.57$ and various couplings down to $(k_{{\rm F}} a_{{\rm F}})^{-1}=1$, and extrapolated eventually the results toward the unitarity
limit, obtaining for this ratio about $0.5$ in agreement with the experimental value.







\begin{thebibliography}{99}


\bibitem{Ketterle-d-i} M.W. Zwierlein, A. Schirotzek, C.H. Schunck, and W. Ketterle, cond-mat/0511197.

\bibitem{Hulet-d-i} G.B. Partridge, W. Li, R.I. Kamar, Y-an Liao, and R.G. Hulet, cond-mat/0511752 (to appear in Science Express, 23
December 2005).

\bibitem{BCS-d-i-homogeneous} R. Combescot, Europhys. Lett. {\bf 55}, 150 (2001);
                              W.V. Liu and F. Wilczek, Phys. Rev. Lett. {\bf 90}, 047002 (2003);
                              H. Caldas, Phys. Rev. A {\bf 69}, 063602 (2004);
                              A. Sedrakian, J. Mur-Petit, A. Polls, and H. M\"{u}ther, Phys. Rev. A {\bf 72}, 013613 (2005).

\bibitem{BCS-d-i-trapped} T. Mizushima, K. Machida, and M. Ichioka, Phys. Rev. Lett. {\bf 94}, 060404 (2005);
                          P. Castorina, M. Grasso, M. Oertel, M. Urban, and D. Zappal\`{a}, Phys. Rev. A {\ 72}, 025601 (2005).

\bibitem{BCS-BEC-d-i-homogeneous} U. Lombardo, P. Nozi\`{e}res, P. Schuck, H.-J. Schulze, and A. Sedrakian,
                                  Phys. Rev. C {\bf 64}, 064314 (2001);
                                  J. Carlson and S. Reddy, Phys. Rev. Lett. {\bf 95}, 060401 (2005);
                                  C.-H. Pao, Shin-Tza Wu, and S.-K. Yip, cond-mat/0506437;
                                  D.T. Son and M.A. Stephanov, cond-mat/0507586;
                                  D.E. Sheehy and L. Radzihovsky, cond-mat/0508430.

\bibitem{PS-2003} P. Pieri and G.C. Strinati, Phys. Rev. Lett. {\bf 91}, 030401 (2003).


\bibitem{STM-1957} G.V. Skorniakov and K.A. Ter-Martirosian, Sov. Phys. JETP {\bf 4}, 648 (1957).

\bibitem{PSS-2004} D.S. Petrov, C. Salomon, and G.V. Shlyapnikov, Phys. Rev. Lett. {\bf 93}, 090404 (2004).

\bibitem{BK-1998} P.F. Bedaque and U. van Kolck, Phys. Lett. B {\bf 428}, 221 (1998).

\bibitem{KC-2005} I.V. Brodsky, A.V. Klaptsov, M.Yu. Kagan, R. Combescot, and X. Leyronas, JETP Letters {\bf 82}, 273 (2005).

\bibitem{Popov} V.N. Popov, \emph{Functional Integrals and Collective Excitations\/} (Cambridge Univ. Press, Cambridge, 1987).

\bibitem{footnote-2} Strictly speaking, in this way one obtains the density of free excess fermions (\ref{free-density}).
                     To recover the density of excess fermions (\ref{interacting-density}) one has to consider, in addition,
                     a self-energy correction of order $\Delta^{2}$ in the $P$-line.

\end{thebibliography}
\end{document}